\documentclass[a4paper,12pt]{article} % DO NOT DELETE THIS LINE
\usepackage{authblk}
\usepackage{blindtext}
\usepackage[english]{babel}
\usepackage{float}
\usepackage{graphicx}
\usepackage{soul}

\begin{document}    % DO NOT DELETE THIS LINE
\date{}

     %-------------------------------------------------------------------------
     % The introductory (header) part of the paper
     %-------------------------------------------------------------------------

     % The title of the paper. Use \shorttitle to indicate an abbreviated title
     % for use in running heads (you will need to uncomment it).

\title{Nanopowder Diffraction}

\author[1]{Zbigniew Kaszkur*}

\author[1]{Ilia Smirnov}

\affil[1]{Institute of Physical Chemistry PAS,Kasprzaka 44/52, 01-224, Warszawa, Poland}

\maketitle                        % DO NOT DELETE THIS LINE

\begin{abstract}

As in the available literature there are still misconceptions about powder diffraction phenomena observed for small nanocrystals ($D<10$ nm), we propose here a systematic and concise review of the involved issues that can be approached by atomistic simulations. Most of phenomenological tools of powder diffraction can be now verified constructing realistic atomistic models, following their thermodynamics and impact on the diffraction pattern. The models concern small cuts of the perfect lattice as well as relaxed nanocrystals also with typical strain and faults, proven by experiments to approximate the real nanocrystals. The discussed examples concern metal nanocrystals. We describe the origin of peak shifts that for cuts of the perfect lattice are mostly due to multiplying of broad profiles by steep slope factors- atomic scattering and Lorentz. The Lorentz factor embedded in the Debye summation is discussed in more detail. For models relaxed with realistic force fields and for the real nanocrystals the peak shift additionally includes contribution from surface relaxation which enables development of an experimental {\it{in situ }} method sensitive to the state of the surface. Such results are briefly reviewed. For small fcc nanocrystals we discuss the importance and effect of multiple (111) cross twinning on the peak shift and height. The strain and size effects for the perfect multitwinned clusters - decahedra and icosahedra, are explained and visualised. The manuscript proposes new methods to interpret powder diffraction patterns of real, defected (twinned) fcc nanoparticles and points to unsuitability of the Rietveld method in this case.

\end{abstract}

     %-------------------------------------------------------------------------
     % The main body of the paper
     %-------------------------------------------------------------------------
     % Now enter the text of the document in multiple \section's, \subsection's
     % and \subsubsection's as required.

\section{Introduction}
The description of diffraction phenomena by the Debye formula and by crystallographic formulas accounting for analytic peak shape (size and strain dependent), peak position determined by Bragg's law and peak intensity related to multiplicity, structure factor, Debye-Waller factor etc., should be equivalent,  at least for larger crystals consisting of many unit cells. When tested on model nanocrystals (NC) the problems arise from complex peak shape often involving side bands and peak shifts departing from the ideal positions determined by Bragg's law. Substantial broadening affects peaks overlap often hindering classic analysis.
The phenomena described by the Debye formula can be observed experimentally. The side bands can be measured for NCs of very narrow size distribution like the reported by Gelisio and Scardi case of 200 peak for cube shaped NCs \cite{Gelisio}. The peak position measured for NCs as a rule deviates from that determined by Bragg's law and different peaks point to a slightly different lattice parameter. This behavior, also easily shown by the Debye formula can be foreseen for broad nanocrystalline peaks even for perfect models (cuts of periodic lattice) remembering that structural peaks in powder diffraction arise from the NCs orientational average, which naturally involves application of the Lorentz factor. The second factor affecting peak position is multiplication of the Debye summation by the average square of atomic scattering factor steeply decaying with scattering angle. 

Discussion of the two approaches` equivalence is given in the next section.

NCs studied experimentally as a rule differ from a cut of a perfect periodic lattice. Either the surface unsaturated bonds exert inward force known as Laplace pressure or these bonds are saturated by chemisorption changing overall force balance. In both cases the resulting structural change (relaxation or reconstruction) is anisotropic (unlike hydrostatic pressure) and affects various diffraction peaks differently. This effect for wide classes of crystals can be effectively modeled. The required models have to describe interatomic interactions for NCs of practical interest- typically larger than 2nm. The number of atoms exceeds then 100 by far and makes quantum mechanical or DFT approaches computationally not feasible. However there exist a number of semi-empirical model interaction potentials that, for metals, can provide reliable relaxed models and reliably describe dynamics of model structures. Such an approach can also provide low energy models of metal clusters deviating from periodicity and forming multiply twinned, defected or strained forms. Discussion of their application in interpretation of experimental phenomena is given in the further sections.

Interpretation of a subtle evolution of peak position, intensity and width measured {\it{in situ}} in response to changing environment can be developed into technique that operates within so called pressure gap and material gap, referring to pressures close to atmospheric and to NCs, hardly approachable by surface science techniques, but able to provide 'surface science' like insight into structure dynamics. The peak position evaluation in powder diffraction is prone to systematic errors limiting its accuracy, however precision and repeatability of such evaluation in series of measurements can be markedly better and depend mostly on counting statistics. For large peak intensity it can easily be within $10^{-4}$ degree of scattering angle ($2 \theta$)(see fig.1). A short review of such results is given in the 'literature examples' section, highlighting the potential of the method. 

In the final part we discuss the diffraction phenomena connected with a local twinning, likely to occur during quick crystallization e.g. in chemical reduction from a liquid phase. We analyze and visualize the effects of multitwinning and strain for the perfect models of decahedra and icosahedra, to extend them to realistic models of clusters randomly twinned in 3D, possible to construct via simulations. The construction method draws attention to the role of vacancies at the initial stages of crystallization from the seed. These ideas are currently developed by us and prepared for publication. The proposed general discussion is supposed to set the ground for the prepared practical methods to analyze diffraction patterns of nanocrystals.

The paper is aimed at filling the gap in common understanding of nanocrystalline diffraction phenomena \cite{Xiong1, Kaszkur4, Xiong2}, promoting and reviewing nanocrystalline {\it{in situ}} techniques as a unique tool to study surface and overall structure evolution of metal NCs, as well as revising interpretation of diffraction patterns of multitwinned structures.

\section{Powder diffraction analysis of perfect nanocrystals - cuts of the perfect lattice.}
A simple illustration of the equivalence of the two approaches to diffraction (crystallographic and Debye's) is peak analysis of powder pattern calculated via the Debye formula for a cluster of atoms cut from a perfect periodic lattice. Knowing the precise (x,y,z) coordinates of all atoms it is straightforward to calculate a histogram of distances (radial distribution) and do the Debye summation. All calculations were performed using the program {\it{Cluster}} with its recent version \cite{Kaszkur8} available to download.
The commonly accepted mathematical model describes the complex amplitude of the scattered wave as:
\begin{equation}
 A(\vec{k}) = \sum_{n=1}^{N} F_{n}(\vec{k}) exp(i\vec{k}\vec{r_{n}})
\end{equation}
where $\vec{k}$ – scattering vector $\vec{k}=\frac{2\pi}{\lambda}(\vec{s}-\vec{s}_{0})$ ;
$\lambda$ - wavelength; $\vec{s}$ - unit vector of scattered wave; $\vec{s}_{0}$ – unit vector of the primary plane wave and $F_{n}$ is the atomic scattering factor of the n-th atom.
The scattered intensity being the square of the amplitude modulus is then : 
\begin{equation}
 I(\vec{k}) = \sum_{i.j=1}^{N} F_{i}(\vec{k}) \bar{F}_{j}(\vec{k}) exp(i\vec{k}\vec{r_{ij}})
\end{equation} 
$\vec{r}_{ij}=\vec{r}_{i}-\vec{r}_{j}$.
Atomic scattering factors are in most cases treated as spherically symmetric and considered as a real function of the length of scattering vector   
 $k = |\vec{k}|= \frac{4\pi}{\lambda} sin(\theta)$, where  $2\theta$ is the scattering angle, $F(\vec{k})=F(k)$. We neglect here all complexities connected with atomic factors for energies close to absorption edge and with appearance of imaginary contribution to F, as well as problems with significant contributions of directional bonds to atom electron density.
 
If all the distances $\vec{r_{ij}}$ participate in the summation (2) in all possible orientations with equal probability (as justified by the large number of randomly oriented NCs), the resulting expression for the scattered intensity should be proportional to a spherical average of (2).
The term:
\begin{equation}
 \frac{sin(k\xi)}{k\xi} = \frac{1}{4\pi} \int\limits_{0}^{2\pi} \int\limits_{0}^{\pi} exp(ik\xi cos(\vartheta)) sin(\vartheta) d\vartheta d\varphi
\end{equation} 
is an average over all spatial orientations of the wave $exp(i\vec{k}\vec{\xi})$.
Then the averaging of (2) results in an expression known as the Debye formula:

\begin{equation}
 I(k)= \sum_{i,j=1}^{N} F_{i}(k) F_{j}(k) \frac{sin(kr_{ij})}{kr_{ij}}
\end{equation}  

where the summation still runs over all possible vectors $\vec{r}_{ij}$ so involves multiply counted distances.
Performing spatial averaging of the simple wave term first, avoids all problems with averaging of peak intensities over crystal orientations after calculating peak intensity, assuming Laue equations and conditions of powder diffraction. However both approaches (crystallographic and Debye's) stem from the same mathematical model of wave scattering and should be equivalent.

As an example, fig.2 shows a powder diffraction pattern for wavelength equal to 0.05 nm of fcc gold lattice (a=0.408 nm) consisting of atoms laying inside a sphere of radius 3.4 nm (fig.2, red curve). The NC consists of 9693 atoms. The pattern divided by the square of atomic scattering factor and the Lorentz factor is given for comparison (black curve). Its intensity is scaled to have peak height corresponding to multiplicity. The correct multiplicity numbers seen for consecutive peaks proves that the Lorentz factor (for powders $1/(sin^2(\theta)cos(\theta)) $ ) \cite{He} is here applicable and buried within the Debye summation formula.
Small differences in the same multiplicity peaks heights are related to varying average lengths of ordered rows of atoms in each crystallographic direction, corresponding to roughness of the surface of the crystal spheroid model. This affects peak width and height but not the intensity.

The peaks were fitted to Voigt functions \cite{Wojdyr1, Wojdyr2}, their positions, intensities and integral widths used to construct:(a) Williamson-Hall plot (fig.4),(b) the figure illustrating the spread of the lattice parameters resulting from each peak position (fig.3) and (c) estimation of Debye-Waller factor by plotting the logarithm of intensities divided by atomic factor, multiplicity and Lorentz factor (fig.5).
It is clear that the broad nanocrystalline peaks are shifted to lower angles when multiplied by steeply decreasing functions like the square of the atomic factor or the Lorentz factor \cite{He} having minimum at about 109 degrees ($2\theta$). This corresponds to increase of the estimated apparent lattice parameter from the lower angle peaks (see fig.3). 

The Lorentz factor for standard powder diffraction traditionally is described as consisting of three contributions: (1) from the range of acceptable misorientation of the crystal, depending on Bragg angle like $1/sin(2\theta)$, (2) from the number of the reflecting crystal planes oriented near Bragg angle (changing like $cos(\theta)$), and (3) from the intensity spread over the circle of $sin(2\theta)$ radius, whereas the detector collects a constant part of it (dependency $1/sin(2\theta)$). Gathering together those multiplicative corrections gives the full Lorentz factor as above. The above description defines its components in terms of contributing crystals whereas the Debye formula can describe intensity from noncrystalline matter, where the used terms of 'misorientation' or 'crystal faces' are not relevant.

It is clear that all these dependencies come from summing the scattered waves and averaging them over all possible orientations of the crystal. The Debye summation formula does the same job but in opposite order- first spherically averaging the phase term of a single wave. The Lorentz factor is applied here alone, without the polarization correction as the state of polarization is not considered in the Debye summation formula.

Fitting of the pattern corrected for atomic scattering factor and Lorentz factor, visibly corrects the lattice parameter from the low angle peaks. Its value from the first peaks is visibly underestimated due to the peak asymmetry. Considering center of the peak mass instead of fit, helps for non-overlapping peaks. Further peaks pose problems with the intensity separation and with the effect of the side bands (ripples). For small NCs the applicability of Bragg's law, even after correction for atomic and Lorentz factor, is thus only approximate. The precise peak positions even for a cut of a perfect lattice depend on its size, shape and type of atoms. However Nelson-Riley lattice parameter extrapolation gives correct value with reasonable accuracy (0.407969 nm vs. 0.408004 nm before correction). Strain estimated from the simple Williamson-Hall plot is close to zero within statistical error (microstrain parameter $\epsilon$=0.0005(4)) (fig.4) as well as a Debye-Waller parameter corresponding to the slope of regression line in fig.5 (slope 0.006(45) with standard deviation of the last figure in parenthesis). Any change of the Lorentz function form results in a visible, evident deviation from the straight line so fig.5 may serve as a simple test of correctness of its analytical formula. Our choice of Voigt functions gives slightly better fit, but the use of other peak shapes (e.g. Pearson VII) provides qualitatively similar results with striking nonlinearity of the plot for the Lorentz factor different than the literature value \cite{He}. Fig.5 also tests significance of the term  $cos(\theta)$ within the full Lorentz factor expression (comparison of black and red curves) that was questioned in \cite{Ozturk}. The authors notice that the intensity may not be proportional to the number of contributing crystals as some crystals can reflect two or more spots into the Debye ring, suggesting correction to the (2) contribution to Lorentz factor, originally formulated as the 'number of reflecting crystals'. However, if a [hkl] poles are equally distributed over the surrounding sphere, then the other, symmetry related, poles (corresponding to multiplicity) in determined orientation in respect to [hkl], are also equally, randomly distributed, resulting in constant density of poles over the sphere (determined by the multiplicity). The correlated sum of random distributions is still a random distribution. Those poles that fulfill Bragg's condition are selected only by the reflection band on the sphere independently of the fact how many crystals contribute to the band. This gives $cos(\theta)$ angle dependence of the number of [hkl] nodes and the (2) contribution to Lorentz factor is still correct.

The Williamson-Hall plot (fig.4) assuming the Scherrer constant to be equal 1 provides an underestimation of crystal size (5.6 nm vs. 6.8 sphere diameter), so the Scherrer constant applicable here is 0.82, differing slightly from the theoretical value calculated for spherical particles (0.89).

The scatter of points in figures 3-5 is due to the Debye formula calculated peak shape that can noticeably differ from the available analytic fitting functions, which results in a misfit. It was discussed by us in the example of a powder diffraction pattern of cubooctahedron (CUB) and fitting with Voigt functions \cite{Kaszkur10}. The Voigt functions being a convolution of Cauchy and Gauss functions have the unique property of having peak profile Fourier coefficients easily expressed analytically \cite{Balzar}. If strain is negligible (like for the discussed perfect model) then those Fourier coefficients can be directly recalculated to a Column Length Distribution (CLD) \cite{Balzar}. For the simple NC model CUB, the CLD in various crystallographic directions can be determined directly from the model (see fig.6 for [111], [200] and [220]). It is clear that the analytic form of CLD differs markedly from that determined directly from the model resulting in the misfit of the peak profile. It shows that for certain perfect NC models the fit of diffraction peaks to some regular analytic profile is never perfect resulting in the error of the estimated peak parameters like position, width and intensity. The significant scatter of points in figures 3-5 is thus for the most part an intrinsic feature of the model and not an error of the fitting routine. For broad overlapping peaks and narrow size distribution the situation is worsened due to the side bands. However a significant improvement in the fit quality can be achieved for a broad distribution of crystal size and shape.

The degree of the peak shift depends on NC size and shape differently for different peaks. An example size dependence for Debye formula calculated patterns of Pd CUBs was given in \cite{Kaszkur3}.

\section{Powder diffraction analysis of real nanocrystals - relaxation, twinning, strain.}
To perform the Debye summation for a model of a real nanocrystal, one has to assume all forces it is subjected to, including interatomic forces. Then, all the atom positions should approach the nanocrystal energy minimum, typically following the gradient of energy vs atom positions (usually conjugated gradient). The final atom coordinates correspond to a local energy minimum.

To effectively simulate forces and cohesive properties of NPs consisting of many thousands of atoms the {\it{ab initio}} quantum mechanical methods become too complex and one has to resort to semi-empirical, classic potentials.
For transition metals the interatomic interactions can be described on the basis of the second moment approximation to tight binding scheme (SMATB) expressing interatomic potentials as n-body interactions. They typically consist of binary repulsive forces and n-body attractive term related to the local electron density within d-band of a metal. N-body interactions cannot be represented by a sum of pair interactions, and are necessary to reproduce basic features of metals, like elastic constants, vacancy formation energy, stacking fault energies, surface structure and relaxation. There is a range of popular potentials providing usually qualitatively similar results. The most general is the Embedded Atom Model potential (EAM)\cite{Daw} employing as the attractive term a so called embedded function that has to be calculated for a given material. Other potentials have various analytic forms e.g. Sutton-Chen potentials \cite{Sutton} are expressed by a sum of power function of interatomic distances as the repulsive term and a square root of the sum of power functions as the attractive term. Other analytic forms are used by Gupta potentials \cite{Cleri} and potentials proposed by Rosato, Guillope and Legrand (known as RGL potentials) \cite{Rosato} using exponential functions of distances. Those different forms of interaction employ numbers of parameters that are tuned to properly describe equilibrium distances, cohesive energy, elastic constants etc. of the material single crystal. They can be also tuned to results of DFT calculations for small fragments of the material. The potentials are mostly proved to be transferable e.g. applicable to atomic clusters different from that they were parametrized for and the parameters are collected in a NIST repository \cite{Hale}. Their accuracy often exceeds that achievable by a simplified quantum mechanical methods (like DFT). 

The atomistic simulations typically encompass energy minimization, molecular dynamics and Monte Carlo methods. Some differences between the potentials appear when calculating surface properties, like surface energy \cite{Tyson} or melting point (via Molcular Dynamics). However the resulting low temperature equilibrium structures do not differ significantly within this class of model potentials and single-crystal surface contraction agrees well with the experimental data \cite{Barnes}. The agreement is good also for NCs if we exclude very small crystals for which the metallic state turns into insulator, the d-band decomposes into discreet electronic levels (e.g. data for dimers) and the model approximation does not work. Otherwise the energy relaxations with those potentials show e.g. stability islands (minimum energy) of various morphologies of fcc metals like icosahedral (ICO), decahedral (DEC) or CUB, that can be verified in XRD or TEM experiments \cite{Cleveland, Barnard, Wells}. Such models are computationally tractable even for larger than 10 nm nanocrystals and numbers of atoms exceeding 50000. 

The effect of surface relaxation can be approached computationally e.g. by considering a series of CUBs growing from a central atom by adding the next (n-th) shell of neighbor atoms. This creates a sequence of so called magic number clusters consisting of N= 13, 55, 147, 309, 561 ... ( N=1+n(11+5n(3+2n))/3 ) atoms. The effect of relaxation can be visualized by comparing peak positions calculated via the Debye formula for two models with the same number of atoms: a relaxed n-shell cluster and the interior of a relaxed (n+1)-shell cluster. The figure 7 shows the 111 peak shift between those two models for a range of crystal sizes calculated using Sutton-Chen potentials for Au and Pt (interatomic potentials after \cite{Sutton}). It is clear that this relaxation shift is much larger than that calculated for cuts of the perfect lattice and for a nanocrystal size of 2-4 nm it exceeds that of the perfect lattice by one order of magnitude (compare fig.7 with fig.3). The 111 peak was chosen as the most intense one and the easiest to monitor in experiment. All these data can be easily recalculated using the program {\it{Cluster}} \cite{Kaszkur8}.

The degree of nanocrystal surface relaxation and its dependence on a state of its surface - chemisorption, reconstruction etc. can be accessed by monitoring the XRD peak position. In experiment, for fcc metals, the peak shift typically corresponds to few tenths of angular degree down to a few hundredths of degree (for Cu tube radiation and 111 peak) or less. The scale of the phenomenon can be thus comparable to the absolute measurement error, however measuring this shift {\it{in situ}} with high precision (as shown in fig.1) during a continuous XRD experiment enables direct detection of the change, when most of the diffractometer systematic errors are repeatable. For this kind of measurements the precision is much more important than the absolute error. For measurements with good counting statistics on a laboratory X-ray source, the error of the peak position shift can be easily less than 0.005 degree of scattering angle \cite{Kaszkur6}, but obviously depends on the sample metal loading.

A change of the peak position due to relaxation can be anisotropic and even may have different direction for different diffraction peaks. It can be shown on a model. Fig.8 presents a model 2057 atom cubooctahedron covered by the next shell atoms divided into those covering (111) facets (red) and those covering (100) facets (green). Modifying the potential of interaction between green and yellow (bulk) atoms, and between red and yellow atoms mimics different interactions of the different facets with adsorbate and relaxation results in the peak positions plotted as the resulting apparent lattice parameter. The adsorbate atoms (usually light) have assumed zero scattering factor and do not contribute to the diffraction pattern related to the structure of yellow atoms only. The effect of the peak shift anisotropy is clearly visible although rows of atoms in the [111] direction do not always terminate on a (111) face. This computational experiment shows the possibility to detect adsorbate bonding force from precise peak shift measurements.

\subsection{Literature examples.}
The peak shift due to relaxation has been measured experimentally and compared to simulations e.g. for Pd, Pt and Au \cite{Kaszkur2, Kaszkur10}. It can be done either {\it{in situ}} by reduction of a metal nanoparticle surface with hydrogen and flushing with helium to shift the thermodynamic equilibrium and cause quick hydrogen desorption, then chemisorbing e.g. oxygen at the surface \cite{Kaszkur1}, or by measuring shift of peak components corresponding to different sizes of nanocrystals resulting in the peak asymmetry \cite{Kaszkur10}. For the studied metals, the shift dependence on crystal size agreed surprisingly well with the simulations. It may serve as experimental proof of the model simulations' applicability.

As the relaxation shift is sensitive to the state of the surface it opens up an experimental way to assess the state of a nanocrystal surface {\it{in situ}}.
The figure 9 shows results of the 220 peak decomposition into Voigt functions for a Au nanocrystalline sample redrawn from \cite{Kaszkur10}. Each Voigt component corresponds to a certain crystal size and its position shift fits well to the shift vs size diagram of fig. 7 (see inset of the figure 9). The 220 peak has been chosen for its relative separation and ease of background estimation. For the above studies the simulations employed Sutton-Chen potentials and developed by us program suite {\it{Cluster}} \cite{Kaszkur8}.

The energy relaxation for small nanocrystals causes besides peaks shift, also appearance of strain that for fully relaxed clusters corresponds to zero overall stress. The atoms are shifted from their periodic lattice nodes and a Williamson-Hall plot displays a small slope of the regression line corresponding to the microstrain parameter (fig.10). For a size distribution of moderate width in spite of the peak asymmetry the Williamson-Hall plot using integral peak width can still provide the volume weighted size average and even Warren-Averbach analysis can be successfully applied to a single mode size distribution \cite{Kaszkur7}. Bimodal distributions already cause more serious problems and cannot be satisfactorily recovered \cite{Kaszkur7}. 

Surface atom shifts from the lattice nodes caused by the energy relaxation affects also appearance of the static non-zero Debye-Waller parameter, but for nanocrystals larger than 5 nm this effect is already not measurable (fig.11).

Unlike for cuts of a perfect lattice, real nanocrystals, characterized by diffraction peaks of the same as single crystal Miller indices, show peak positions departing from Bragg's law even after correction for atomic scattering factor and Lorentz factor. The differences are due to energy anisotropic relaxation of different crystal faces. Phenomena of chemisorption and/or surface reconstruction can cause diffraction peak shifts that can differ for various crystallographic directions (peaks) in magnitude and sign as shown in the previous section.

The above results show that atomistic simulations can provide satisfactory explanation of various features of nanocrystalline patterns and that nanopowder diffraction can be used as a surface sensitive method. It can be developed into a method of analysis of surface evolution: chemisorption, relaxation, reconstruction as well as reshaping and segregation in alloys. The simulations are able to show reliable thermodynamic disorder effects, conformational changes or concentration diffusion gradients. Most of these effects are reported in literature independently of the actual potential scheme used (e.g. Gupta or Sutton-Chen) \cite{Cleri, Sutton} and their application in nanopowder diffraction analysis cannot be deprecated as arbitrary and unreliable.

Examples of experimental studies analyzed with help from atomistic simulations and employing surface relaxation effects include {\it{in situ}} diffraction observation of Pt nanocrystal surface reconstruction during desorption of hydrogen \cite{Rzeszotarski}. The process could be monitored due to slowing down of desorption by reverse spill-over of hydrogen from the supporting silica substrate. Another observation concerned quick, reconstruction driven coalescence of Pt nanocrystals in NO atmosphere at 80 deg.C and change of Pt nanocrystal shape in CO atmosphere \cite{Kaszkur6, Kaszkur9}. Atomistic simulations could also help explain different diffusion mechanisms in repeatable segregation of Pd or Ag in PdAg alloy exposed to CO or He atmosphere respectively \cite{Kaszkur5}. 

\subsection{Strain in simple model nanocrystals.}
Energy relaxation of metal nanocrystals constructed as cuts of the perfect fcc lattice results in surface contraction followed in deeper layers with oscillating interplanar distance approaching the bulk value \cite{Barnes}. The shift in average lattice parameter, as was shown above, can be measured experimentally. The strain is localized at the surface and differs for different crystal facets (fig.12 compared to fig.13). As a rule all SMATB potentials point to less symmetric small size clusters as being more stable than fcc crystals. Theoretically ICO shape dominates for the smallest metal nanoclusters, followed by Marks' DEC \cite{Marks} and fcc (CUB). Such clusters are examples of crystals multiply twinned in respect to (111) planes and can be easily modeled and energy relaxed. As in the fcc metals the change in stacking order of (111) planes causes only small energy difference, the appearance of the ABA, hcp-like sequence may be frequent with the mirror plane B. If the crystal contains point defects (e.g. vacancies), a random 3D system of crossing local mirror planes can form a complex structure. The relaxation (e.g. conjugate gradient energy minimization following Fletcher-Reeves algorithm) results in strain located at the surface and close to the twinning planes. As the net forces acting on atoms are zero at minimum energy, this strain corresponds to zero overall stress. 

Strain in powder diffraction is mostly reflected in peak broadening propagating with scattering angle as $tan(\theta ) $ and proportional to a local variation of the interplanar spacing. Accounting for defects e.g. planar, like notable occurrence of polytypic (111) sequences, may lead to peak shifts \cite{Warren}. In three dimensions it can be visualized as a 'volumetric strain' i.e. the magnitude of the atom shift from its periodic lattice position as given by the program {\it{OVITO}} \cite{Stukowski}. This is how the stress was evaluated in fig.12. The shift can be evaluated independently for all fcc domains omitting atoms at (111) twin planes which have hcp-like coordination, where an ABCA sequence of these planes turns into ABA. Alternatively one can visualize local strain mapping cohesive energy of each atom of the cluster and assume, that the difference in energy of neighbor atoms corresponds to a local stress, that results in change of the interatomic distance i.e. strain (see fig.13). The comparison of fig.12 and 13 may serve as graphical illustration of Hooke's law. 

Twinning may be a major factor affecting powder diffraction patterns and the pattern calculated for the one (out of five) sectors (domains) of DEC multiplied by 5 differs from the pattern of the full DEC mostly in intensities of peaks closely related to 111, with some [111] directions being continued across the twin plane, contributing to intensity (fig.14). Strain, localized close to twinning planes, causes peak shifts (e.g. approach of 111 and 200) and peak broadening. With increasing size of DECs, the strain, localized only to twin planes, affects less the peak position, which approaches that of an fcc structure. 

Around each of six 5-fold axes of ICO, one can cut off a sub-crystal by sectioning it by planes connecting neighbor corners with the center of the ICO. Such figure looks like a DEC but the sectioned part (internal in ICO) is subjected to different relaxation forces. Diffraction patterns of model ICOs are thus strain-ridden and fig.15 shows the strain distribution in one of five fcc sectors of such an ICO. Half of such sector shows an area of contraction, the other half, an area of expansion of interatomic distances. The powder diffraction pattern of ICO departs from that of fcc markedly more than in the case of DEC. The 220 peak splits then into two. Again with the [111] directions partly continued across the twin planes, the 111 peak grows. Other peaks with increasing crystal size suffer substantial strain being moved and broadened.

\subsection{Modeling of strain.}
Analyzing nanocrystalline diffraction patterns of metals quite often requires accounting for a small size non-periodic lattice forms like Marks' or Ino DECs and ICOs \cite{Marks}. Fitting such patterns to a sum of perfect forms is always possible having enough fitting parameters \cite{Cervellino}. However the occurrence of such perfect forms alone is thermodynamically unlikely and indeed was questioned in literature with suggestion of more strained forms domination \cite{Longo}. Those strained forms for fcc metals grow typically with tendency to locally disturbed ordering of (111) layers from the perfect ABCA sequence. Having eight [111] directions, in three dimensions it results in a complex multitwinned structure with many locally ordered domains. Such a structure is hard to be described analytically by fitting parameters in a full profile fitting procedure. Instead, Longo \& Martorana \cite{Longo} could satisfactorily fit experimental patterns to a one dimensional statistical model of polytypy describing the diffraction pattern by parameterized Debye summation formula and using twinning probabilities defined by Warren \cite{Warren}. Although the number of used parameters enables a good fit, it again, is thermodynamically unlikely for twinning to happen in one direction only (1D). The most common examples like DECs and ICOs show a natural tendency to 2D and 3D twinning. The characteristic evolution of 111 and 200 peaks shape of DECs with their size (fig.16) can be described by statistical parameters of Longo \& Martorana \cite{Longo}, but evidently the structure encompasses 2D twinned domains with all twinning planes parallel to the [220] 5-fold axis.

It seems that atomistic modelling can provide multitwinned structures in a way mimicking natural random growth process. Our approach assumes random selection of atoms to be deleted, forming vacancies, and relaxing the resulting structure. If the vacancy concentration exceeds a certain level the structure collapses leading to formation of locally ordered domains \cite{Smirnov}. Diffraction patterns of such structures, calculated via the Debye formula, show striking similarity to various models of DECs - the 111 and 200 peaks approach each other with some intensity forming bridge between them. However such multidomain structure can result in more complex diffraction pattern features than for a DEC of similar peakwidth. It may cause even stronger 111 and 200 peak shifts, that approach each other closer than for DEC, and provide a different ratio of peak intensities.  Fig.16 shows the patterns of 3D multidomain structures obtained following our method compared to the DEC patterns. A section of one of the obtained model nanoparticles with domains marked in colors is presented also in fig.15. Many published patterns of nanocrystalline Au or Ag have striking features of such multidomain structures but have been analyzed as fcc structure.  Evidently their 111 peak position would not point to the lattice parameter - difficult to define for multitwinned structures. 

For small metal nanocrystals the multitwinned and/or ICO shape should dominate but it is hard to find experimental examples of the ICO diffraction patterns (fig. 17) although such crystals are observed via TEM.
Larger crystals are preferentially DEC or single domain fcc. For DEC crystals there is one [220] direction, common to all domains so the 220 reflection is less affected by twinning. (It should provide a reasonable Scherrer size and be accessible to peak-shape analysis like Warren-Averbach.)
Interpretation of details of NC diffraction pattern using atomistic simulations opens up the possibility of surface science like studies of NC metals {\it{in situ}}. However any structural conclusions about crystal interior (atoms A) and the surface (atoms B) has to take into consideration the fact that to diffraction pattern contribute distances A-A, B-B and also A-B. If A is described as a crystalline phase then B forms another phase and the contribution from B-B is comparable to that of A-B (see fig.18). However the contribution from A-B distances cannot be included in Rietveld analysis. This excludes such analysis as non-applicable. This applies also to non-periodic quasi-crystalline forms that cannot be described in crystallographic terms. Also the surface relaxation depending on strength of interaction on different facets of the NC may cause peak shifts that differ from peak to peak, even in direction. 

\section{Conclusions}
We show the origin of NC diffraction peak shift with the peak positions departing from Bragg's law. For NCs being cuts of the perfect lattice the shift is mostly caused by broad peaks being multiplied by quickly changing with angle functions - the square of atomic factors and the Lorentz factor. 
We stress the importance of the Lorentz factor that is contained within the Debye summation formula as the formula results from spherical averaging of the phase term. This fact is usually not realized and the peak shifts are often wrongly interpreted in the literature. 

For simulated models of the relaxed NCs the departure from Bragg's law has much more complex character and can provide subtle structural details. We provide data showing good applicability of such models to describe features of the real NCs. The presented data explain sources of non-Bragg features of powder diffraction and illustrate the interpretation of phenomena measurable by {\it{in situ}} powder diffraction on NCs, via atomistic simulations and the Debye summation formula. 

For NCs of size less than 10 nm application of a full profile Rietveld analysis tends to hide subtle surface effects and conceals structural details. We show the potential of our technique as a surface science tool allowing monitoring of the state of surface during physico-chemical processes e.g. chemical reaction. 
Instead of considering the idealized models of NCs like cuts of a perfect lattice or ideal approximants of quasicrystals, this approach enables accounting for complex relaxation phenomena and thermodynamics of the system driven by vacancies and stacking faults, especially for metals at low temperatures after initial preparation by seeding and chemical reduction. The simulations provide a new way to interpret powder diffraction patterns of multidomain random structures. 

All atomistic simulation data are available to download and reproduce using our simulation suite {\it{Cluster}} \cite{Kaszkur8} or can be supplied on request.

     % Appendices appear after the main body of the text. They are prefixed by
     % a single \appendix declaration, and are then structured just like the
     % body text.

     %-------------------------------------------------------------------------
     % The back matter of the paper - acknowledgements and bibitems
     %-------------------------------------------------------------------------

     % Acknowledgements come after the appendices

\section{Acknowledgments}

The work has been financially supported by The Polish National Science Center (NCN) under Research Grant No 2018/29/B/ST4/00710 .

     % References are at the end of the document, between \begin{bibitems}
     % and \end{bibitems} tags. Each bibitem is in a \bibitem entry.

 \begin{figure}
 \centering
 \caption{Peak position measurement statistics for 210 peak of $LaB_{6}$ NIST standard using Cu sealed tube (2.2 kW) charged 
 with 40 kV and 40 mA. It encompasses 250 measurements over approx. 25 hours. The first 1.5 hour data showed anode temperature stabilization and were omitted.The distribution is close to Gaussian and all data fall within $3\sigma$ margin. }
 \includegraphics[width=1\columnwidth]{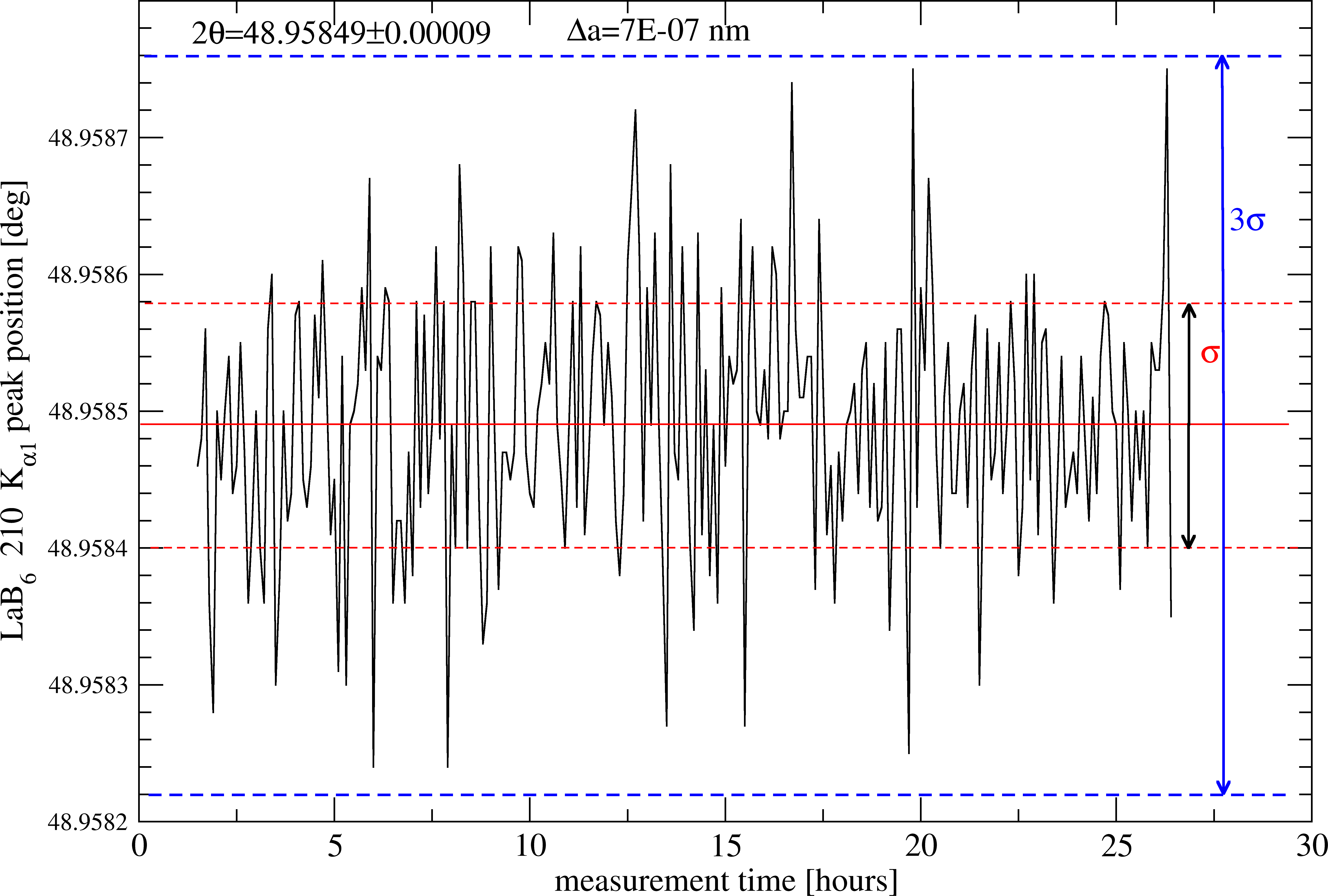}
 % \label{fgr:Fig1}
\end{figure}

  \begin{figure}
 \centering
 \caption{Calculated powder diffraction pattern of a perfect lattice spherical gold nanocrystal (red). The pattern divided by the square of the atomic scattering factor and the Lorentz factor scaled to multiplicity (black).}
 \includegraphics[width=1\columnwidth]{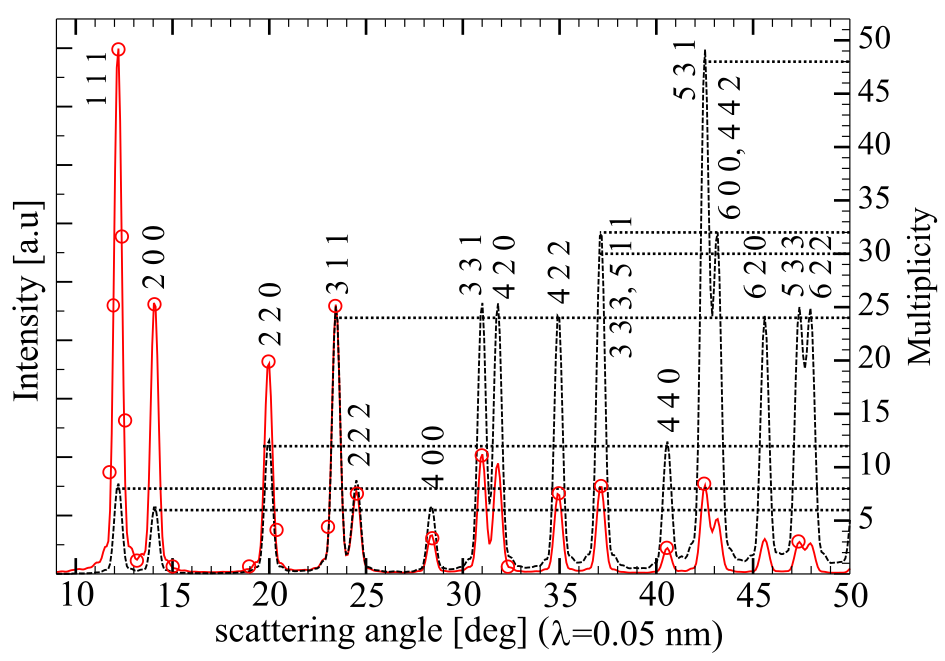}
 % \label{fgr:Fig2}
\end{figure}

\begin{figure}
 \centering
 \caption{Apparent Lattice Parameter (ALP) from Fig.2 peaks fitted to Voigt profiles. Fit to peaks as calculated from the Debye formula (red circles), fit to peaks corrected for the square of atomic factor and the Lorentz factor (black squares).}
 \includegraphics[width=1\columnwidth]{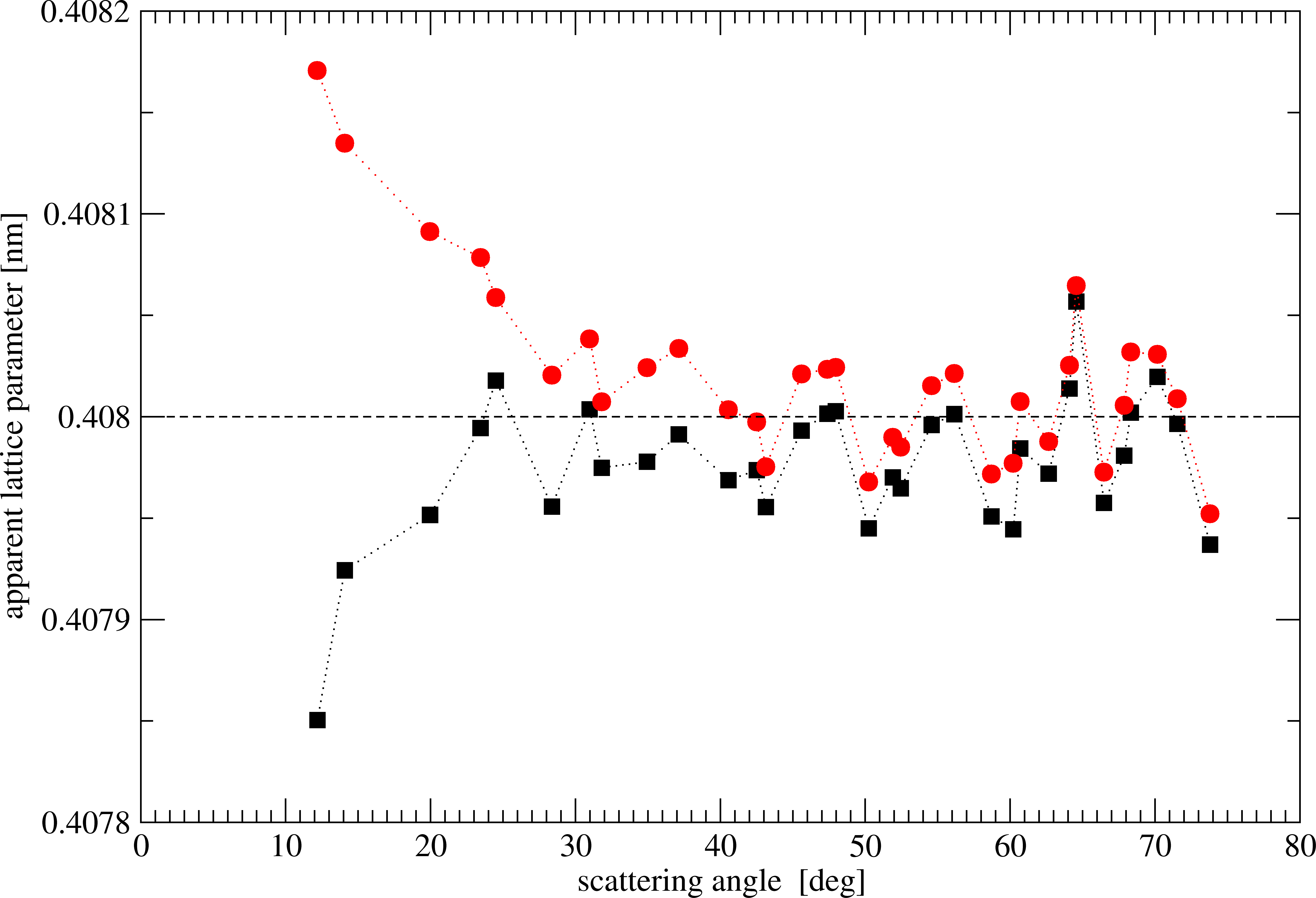}
 % \label{fgr:Fig3}
\end{figure}

 \begin{figure}
 \caption{Williamson-Hall plot for Fig.2 peaks fitted to Voigt profiles. Fit to peaks as calculated from the Debye formula (green circles) gives microstrain parameter $\epsilon$ close to zero (within the estimated error), fit to peaks corrected for the square of atomic factor and the Lorentz factor (black squares) results in $\epsilon$ slightly exceeding the linear regression error. As the model is strain free the values of microstrain parameter are the measure of systematic error in evaluation.}
 \includegraphics[width=1\columnwidth]{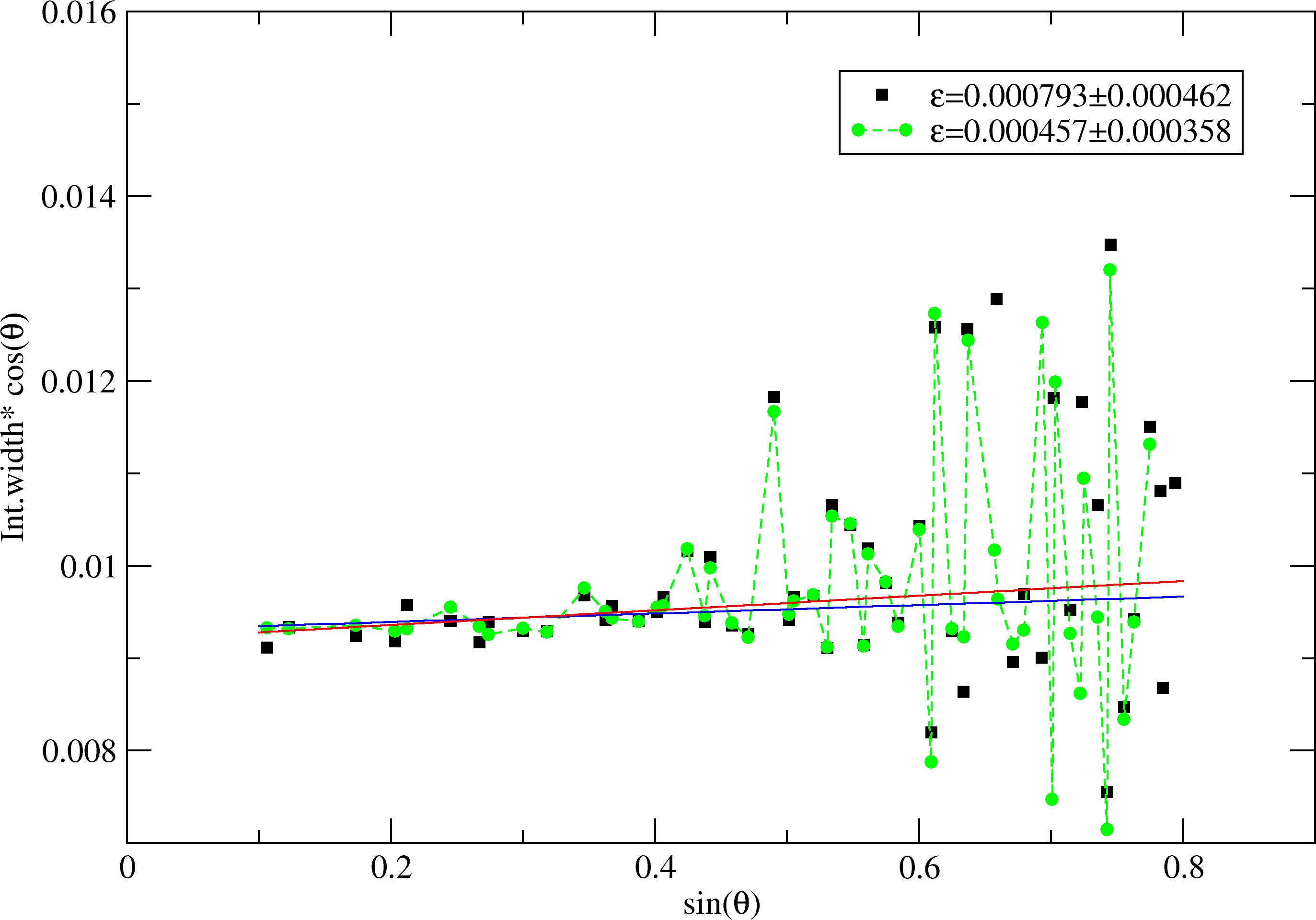}
 % \label{fgr:Fig4}
\end{figure} 

\begin{figure}
 \centering
 \caption{Debye-Waller factor estimation from Fig.2 pattern divided by the square of atomic factor and the Lorentz factor. The peaks were fitted to Voigt profiles and logarithm of peak intensity divided by multiplicity is plotted for the correct Lorentz factor (black points). Any error in the Lorentz factor formula shows marked deviations from linearity, here (blue points) plotted for the Lorentz factor wrongly assumed to be equal to $1/sin(2\theta )$ , and for the Lorentz factor wrongly assumed to be equal to $1/sin^{2}(\theta )/cos^{2}(\theta)$. The figure proves that the Lorentz factor is already included into the Debye formula providing zero slope and good linear regression only for the correct form of the factor. The slope value corresponds to atoms mean square displacements from nodal position - zero for the perfect lattice.}
 \includegraphics[width=1\columnwidth]{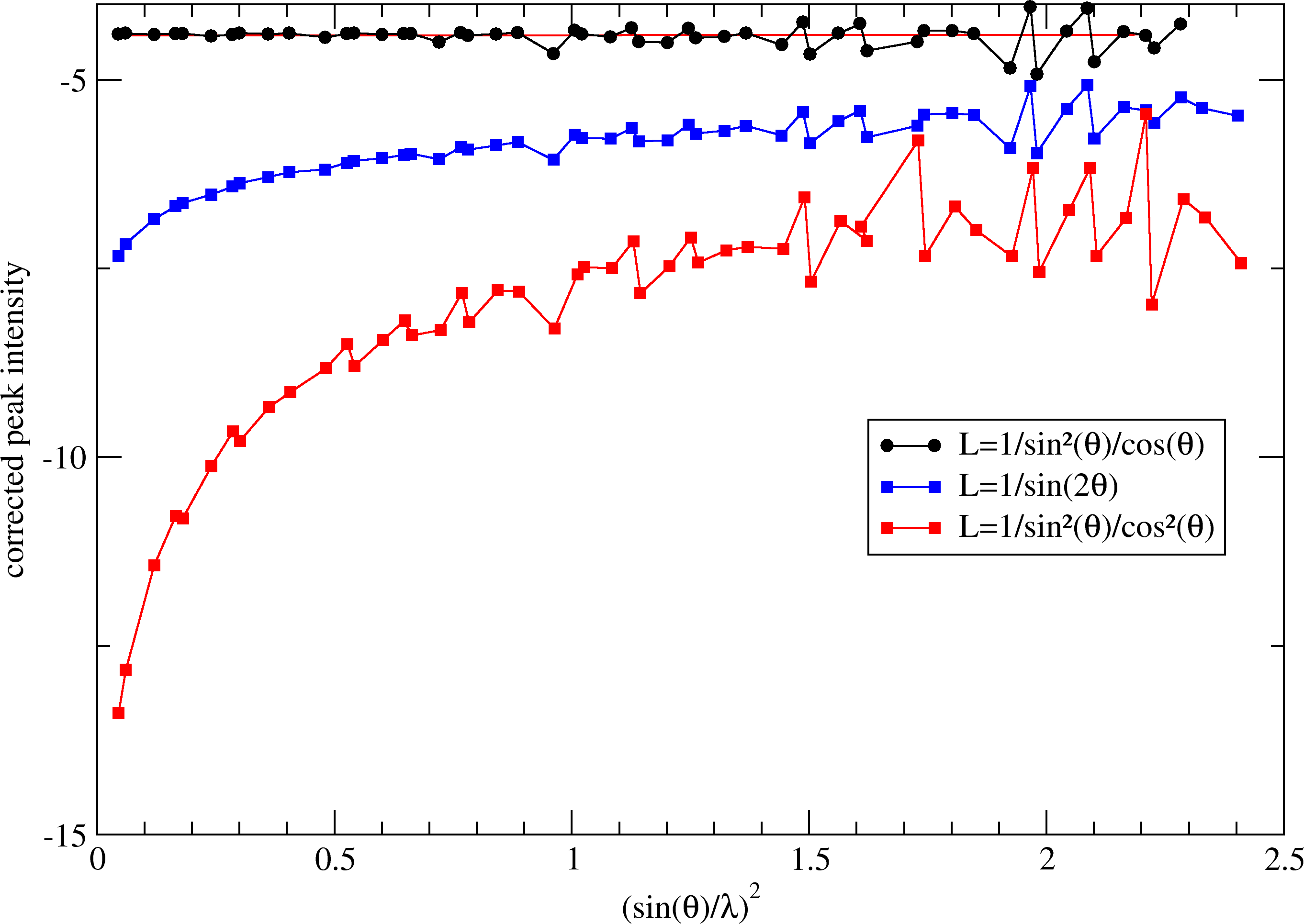}
 % \label{fgr:Fig5}
\end{figure}

\begin{figure}
 \centering
 \caption{Column Length Distribution (CLD) calculated directly from the cubooctahedron model for three peaks 111, 200 and 220, compared to its representation from peak shape analysis when peaks are fitted to a Voigt function (violet, orange and cyan line for various Voigt parameters) and to CLD for the isotropic ball model (green line) (redrawn from \cite{Kaszkur10}). As seen, the monosize cubooctahedron diffraction peaks cannot be well described by a Voigt function, nor by any other regular analytic form. However their broader size distributions can likely be approximated with a Voigt function. For 111 peak the CLD cannot be approximated even assuming a high value of the Cauchy to Gaussian broadenings ratio.}
 \includegraphics[width=0.7\columnwidth]{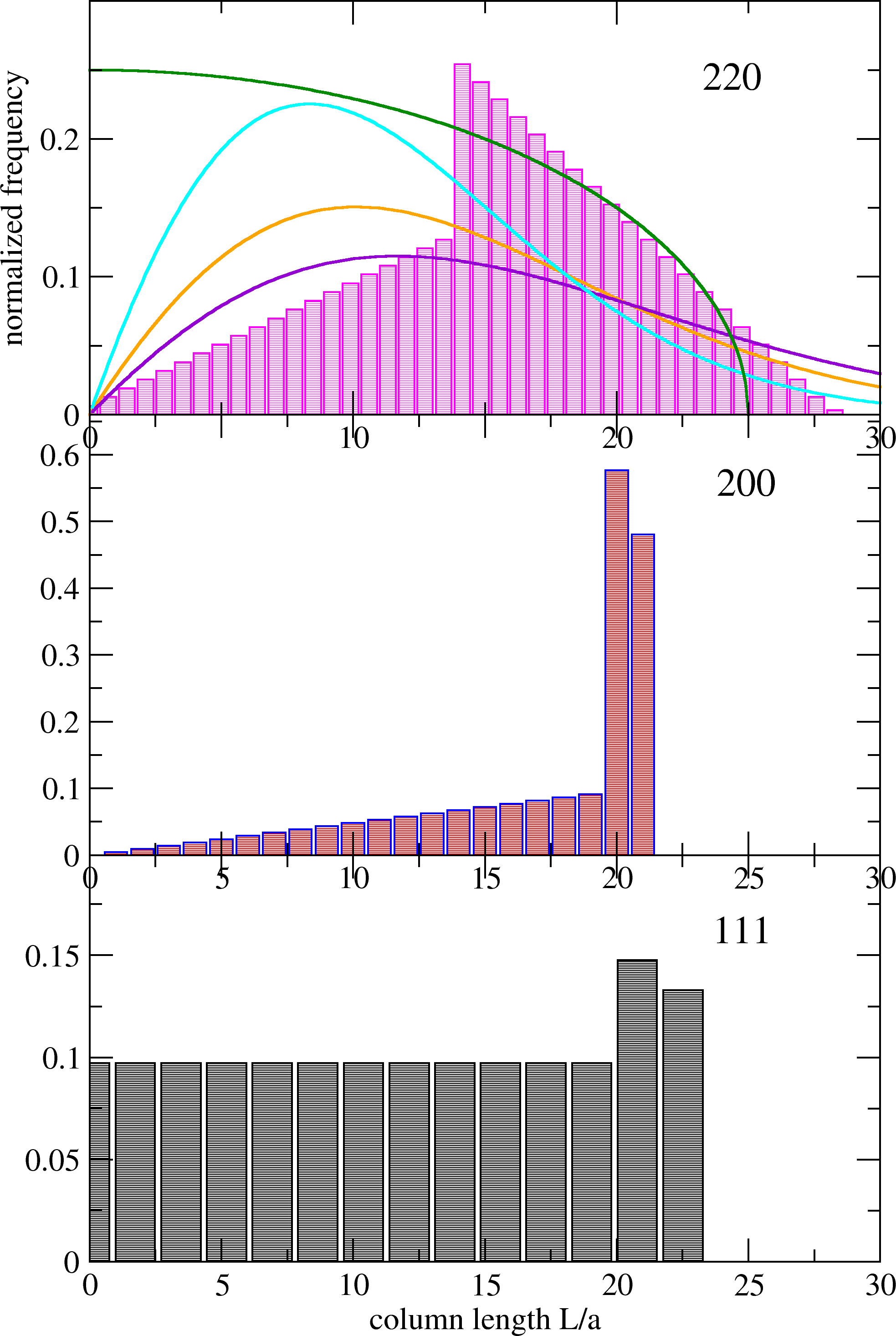}
 % \label{fgr:Fig6}
\end{figure}

\begin{figure}
 \centering
 \caption{Calculated Apparent Lattice Parameter (ALP) shift due to surface relaxation (contraction) for Au and Pt versus crystal size. Marked are experimentally measured shifts caused by chemisorption of several gases on Pt (redrawn from \cite{Kaszkur10}).}
 \includegraphics[width=0.8\columnwidth]{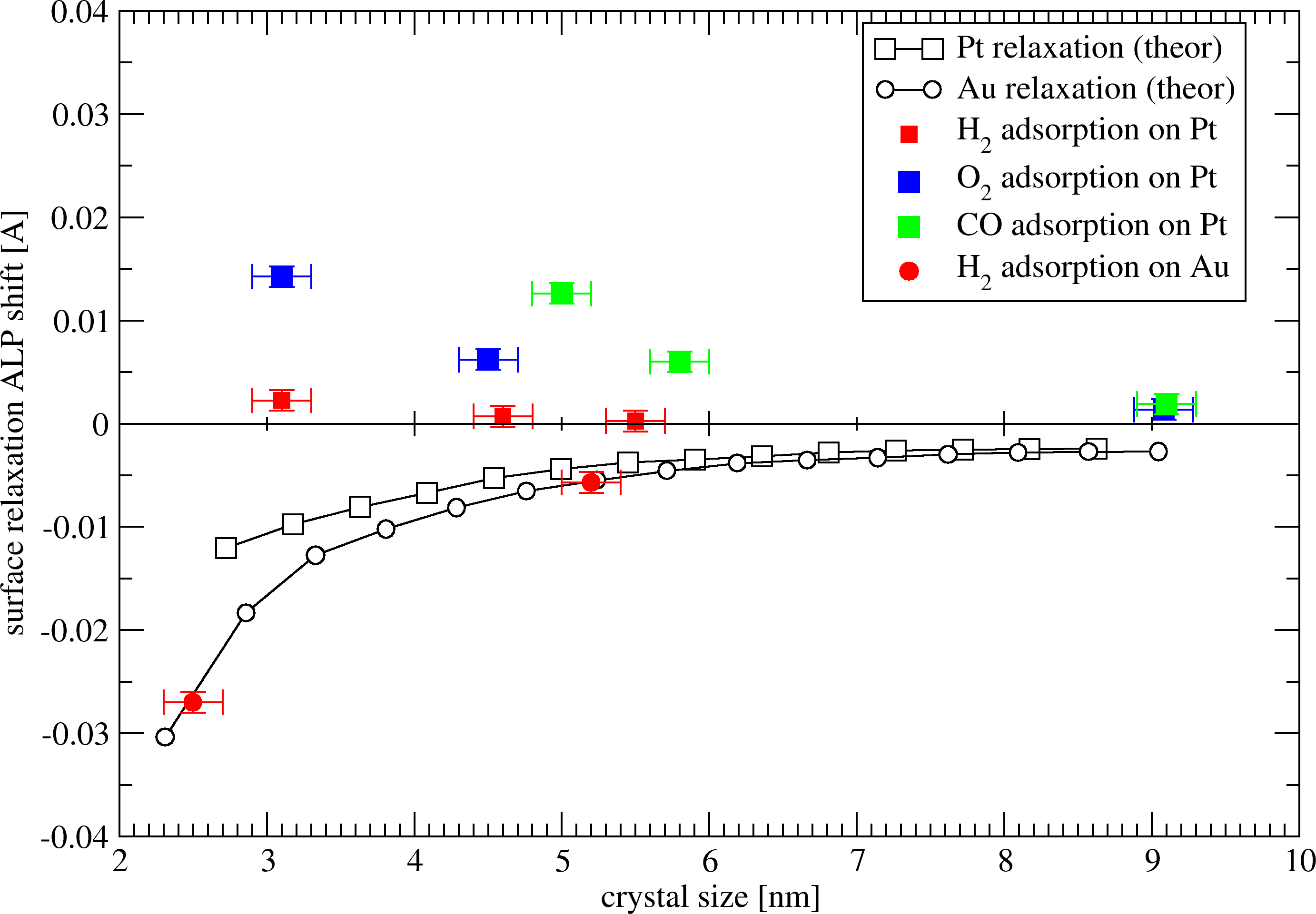}
 % \label{fgr:Fig7}
\end{figure}

\begin{figure}
 \centering
 \caption{Calculated Apparent Lattice Parameter (ALP in Angstrom) for a range of diffraction peaks and different interaction forces with a model adsorbate on (111) (red atoms) and (100) (green atoms) faces. Strength of interaction with (111) plane is increasing and with (100) decreasing in order of lines: blue, yellow, red, green.}
 \includegraphics[width=0.8\columnwidth]{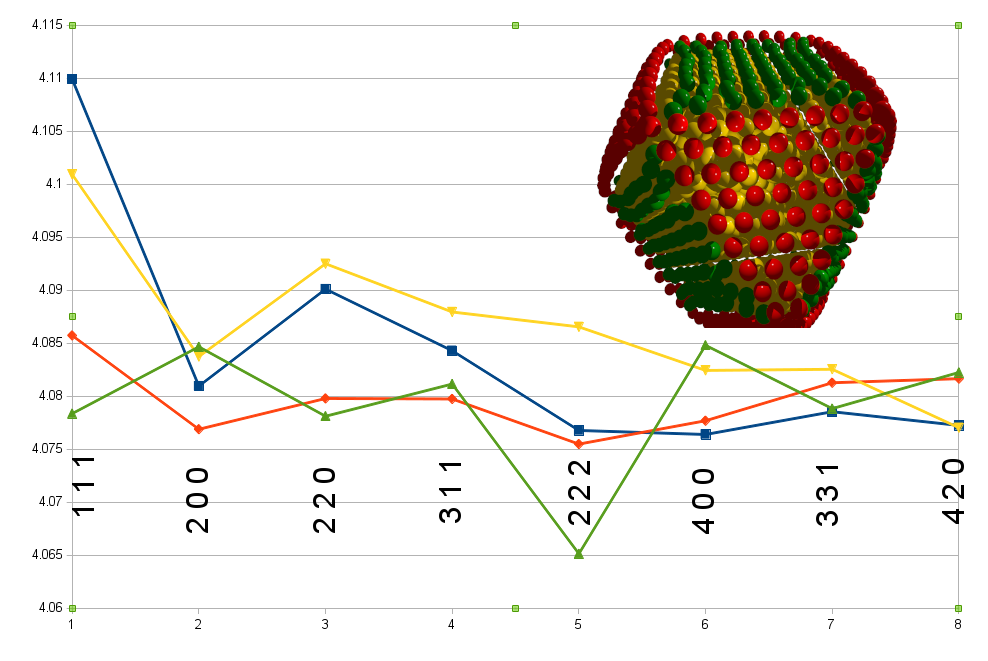}
 % \label{fgr:Fig8}
\end{figure}

\begin{figure}
 \centering
 \caption{Decomposition of the 220 peak of nanocrystalline Au in He, $O_2$ and $H_2$ atmospheres onto several Voigt profiles (redrawn from \cite{Kaszkur10}). The component peak shifts are representative of the Apparent Lattice Parameter (ALP) shift, the peak widths represent the component crystal size in the [220] direction. Inset plot follows well the ALP plot for Au in fig.7.}
 \includegraphics[width=0.8\columnwidth]{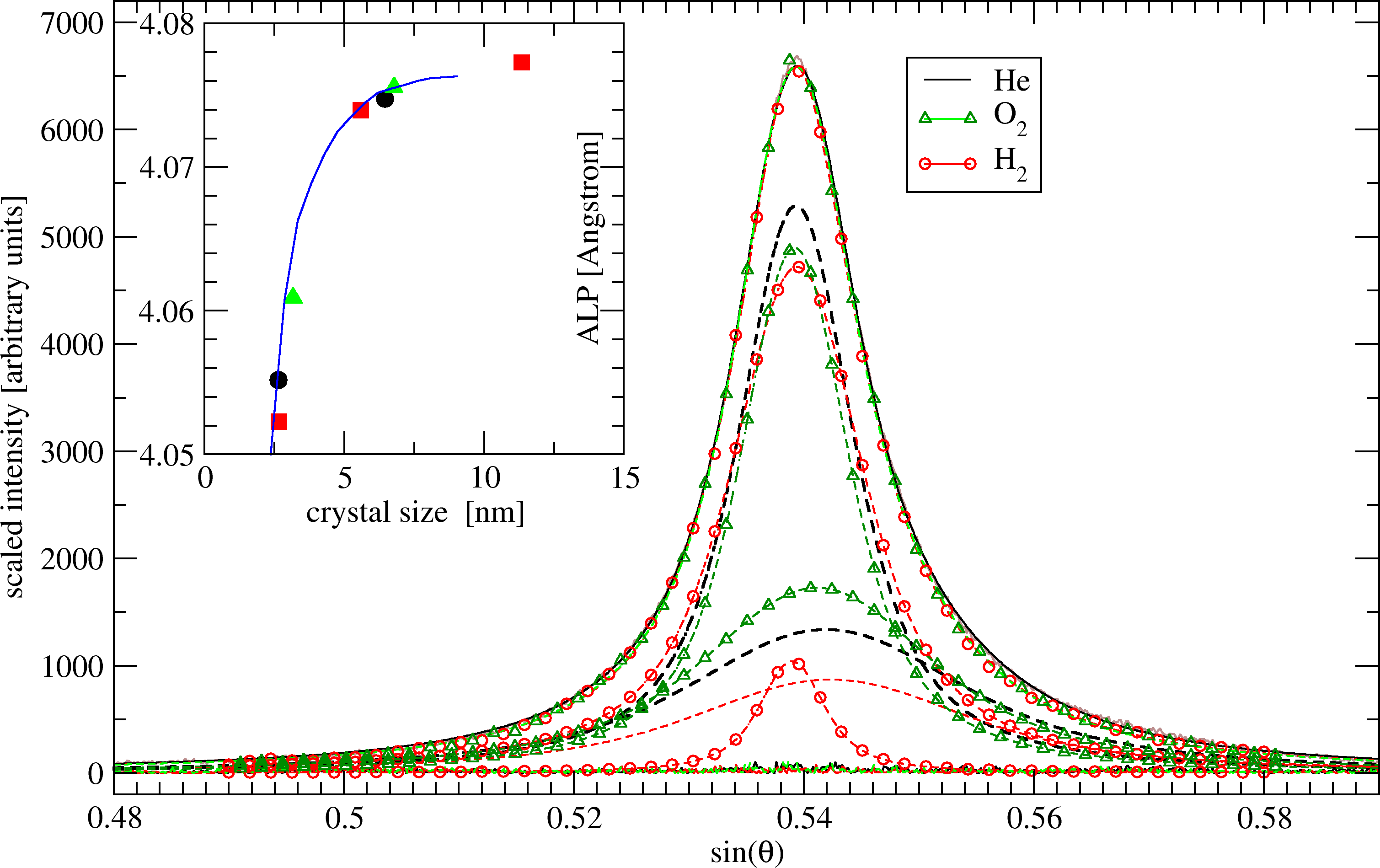}
 % \label{fgr:Fig9}
\end{figure}

\begin{figure}
 \centering
 \caption{Williamson-Hall plots for the calculated patterns of 9693 atom spherical Au cluster. The plots are for the cut of perfect lattice (red squares) and for the relaxed cluster (black circles).}
 \includegraphics[width=0.8\columnwidth]{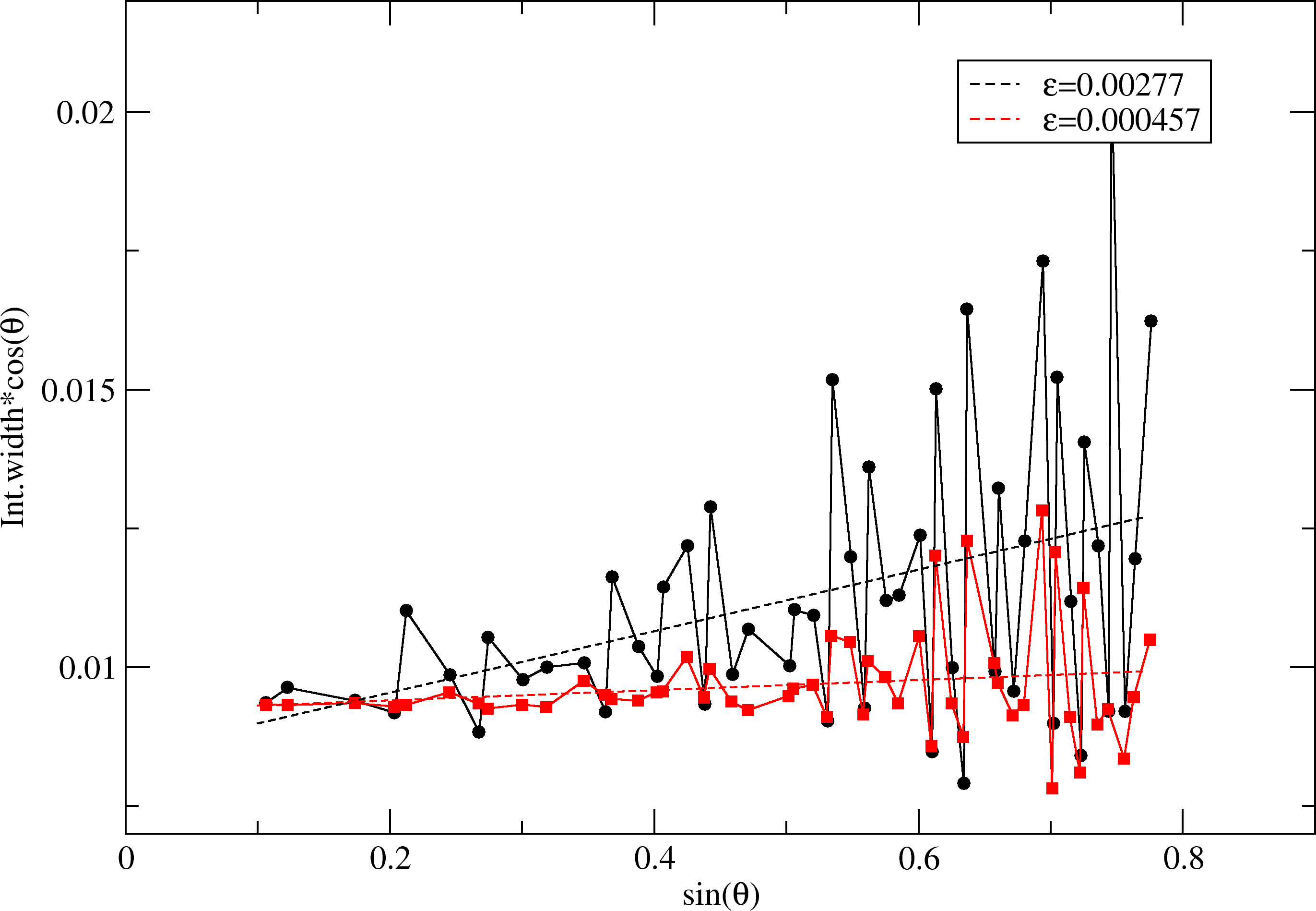}
 % \label{fgr:Fig10}
\end{figure}

\begin{figure}
 \centering
 \caption{Logarithm of the peak integral intensity of 9693 atom spherical Au cluster, corrected for the square of atomic factor, the Lorentz factor and muliplicity. The plot is for the cut of the perfect lattice (red squares) and for the relaxed cluster (black circles). In both cases the slope, proportional to Debye-Waller parameter, is zero within the linear regression error.}
 \includegraphics[width=0.8\columnwidth]{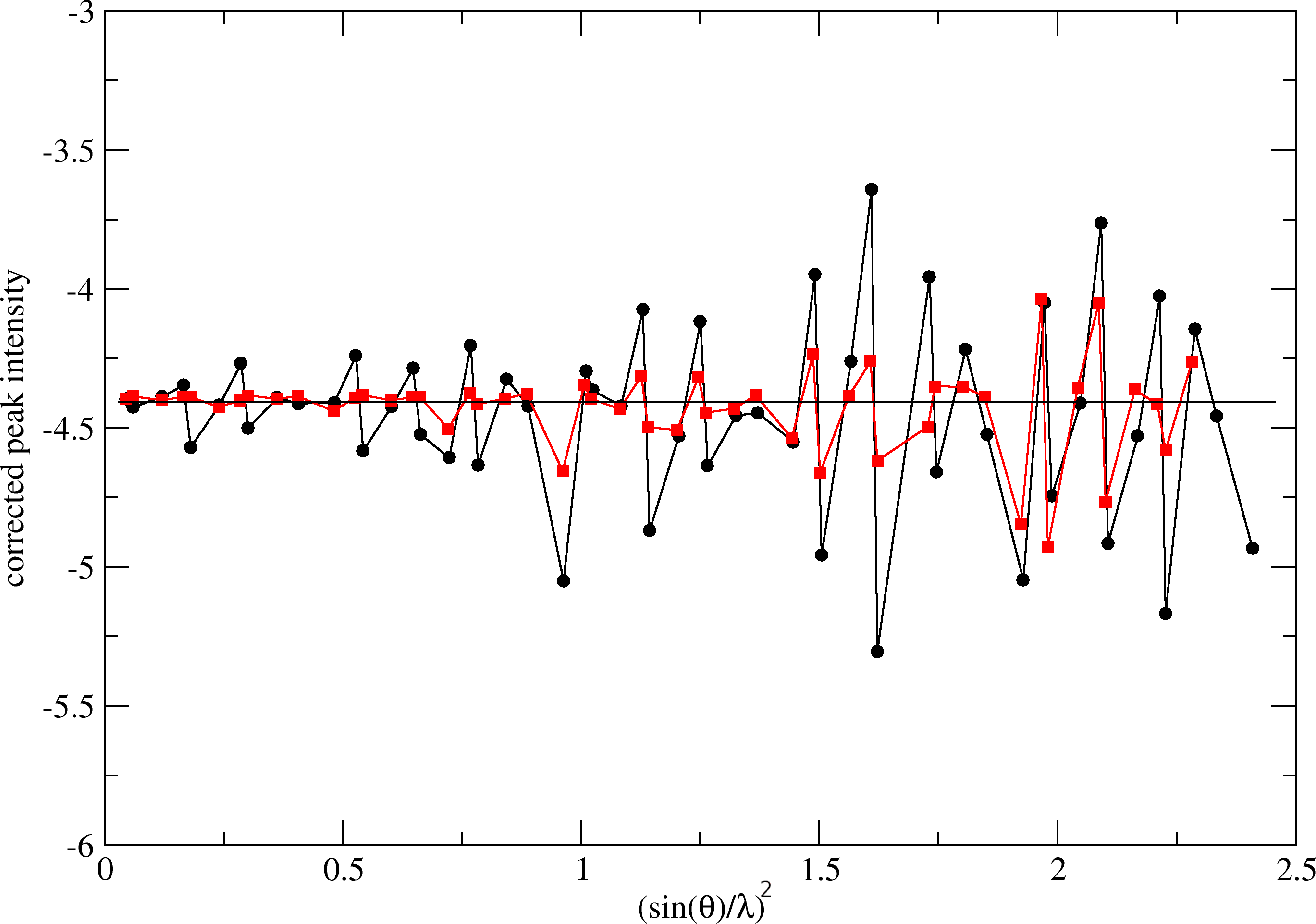}
 % \label{fgr:Fig11}
\end{figure}

\begin{figure}
 \centering
 \caption{Illustration of 'volumetric' strain as given by the {\it{Ovito}} program \cite{Stukowski} for a) a relaxed cubooctahedron (CUB), b) a relaxed Marks' decahedron (DEC), and c) a relaxed icosahedron (ICO), all comprising close to 1985 atoms. The sections shown are through the center perpendicular to the [220] direction. The color scale corresponds (in percents) to a ratio of the atom shift from the perfect atom position (within a given sector) to the perfect lattice interatomic distance. The plot d) shows the radial distribution of strain within the considered three models. }
 \includegraphics[width=0.8\columnwidth]{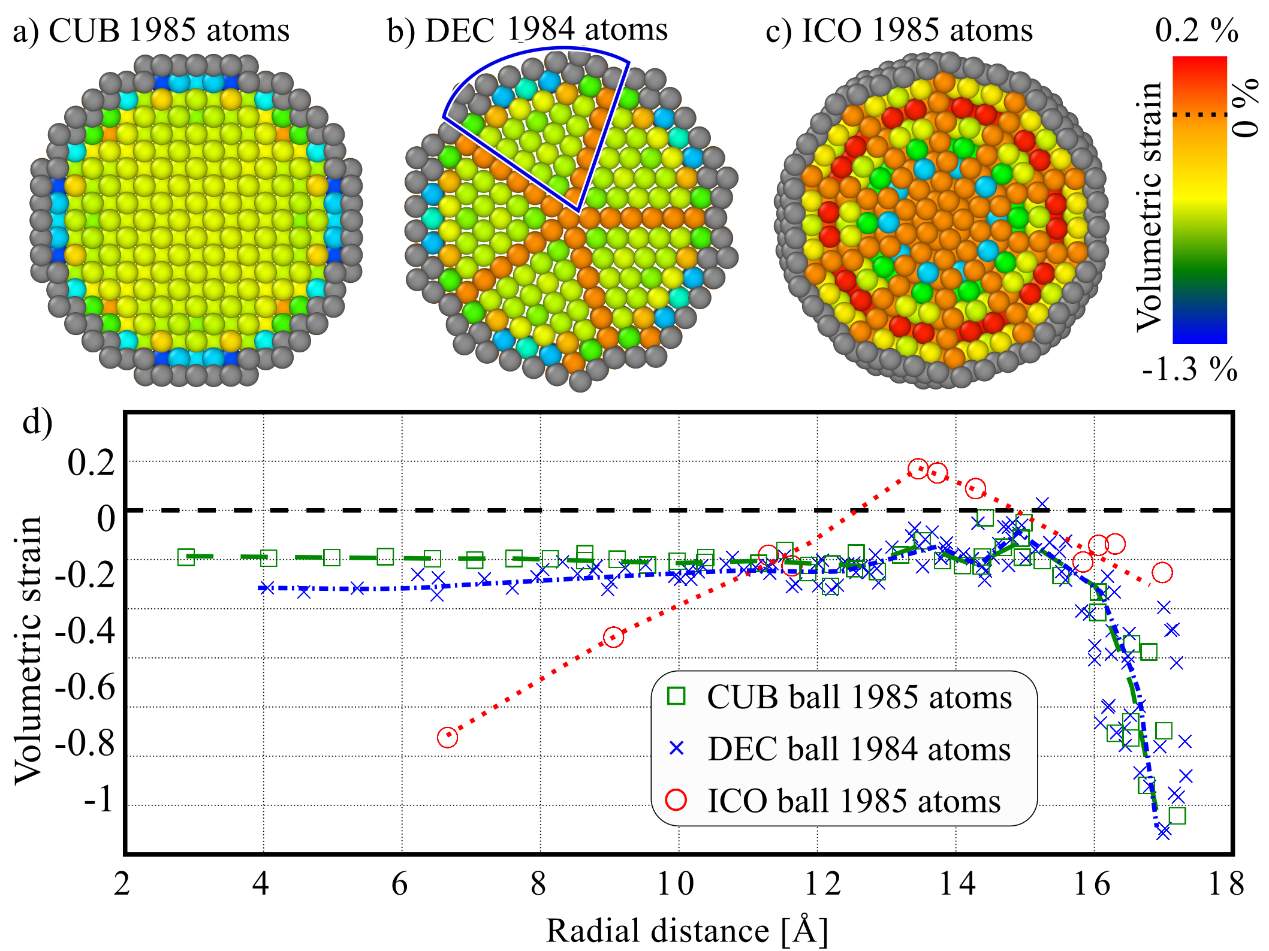}
 % \label{fgr:Fig12}
\end{figure}

\begin{figure}
 \centering
 \caption{Map of atom cohesive energy in sections of a cubooctahedron (first column), a Marks' decahedron (second column) and a icosahedron (third column).The sections a) to e) present the color map of cohesive (potential) energy in different scales marked by the color bar, visualizing  energy differences in deeper layers of the clusters.}
 \includegraphics[width=0.8\columnwidth]{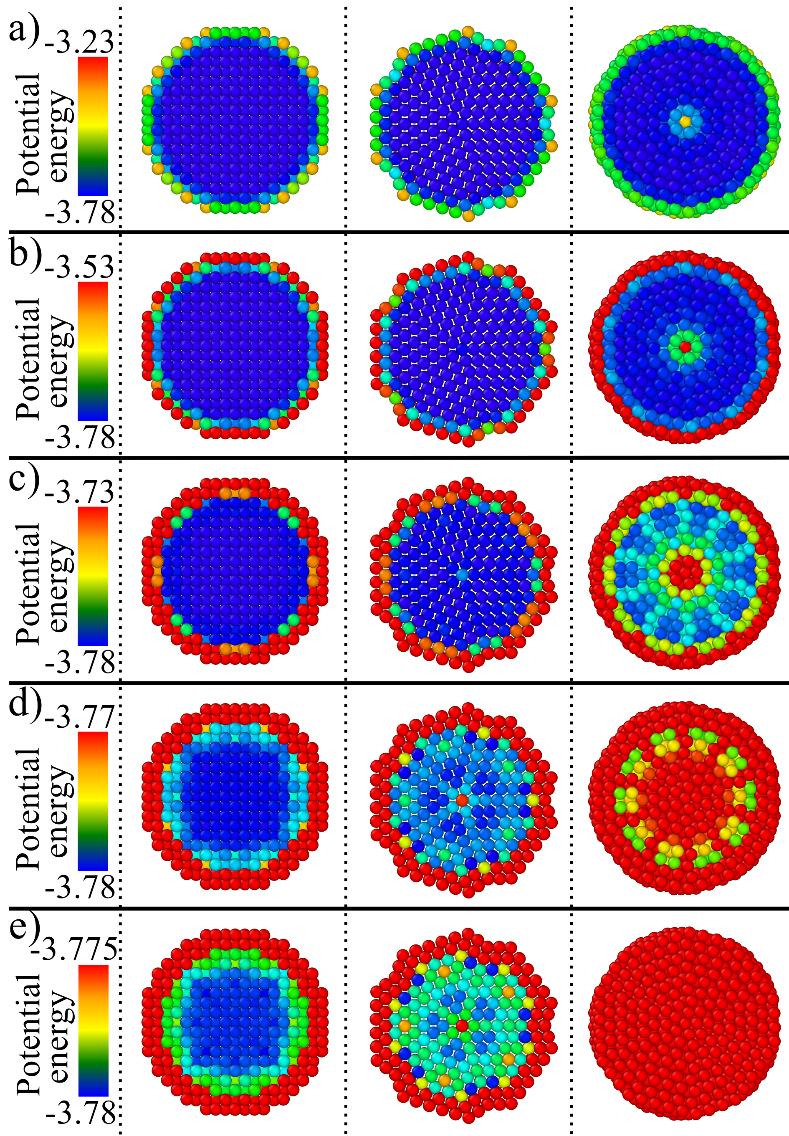}
 % \label{fgr:Fig13}
\end{figure}

\begin{figure}
 \centering
 \caption{Diffraction pattern of the relaxed decahedron (as for Cu $K_{\alpha}$ radiation) compared to the pattern of one (out of five) sector (domain) multiplied by 5. It shows that the effect of strain in the single domain, determining approach of 111 and 200 peak positions, is the same as for the whole decahedron. Correlations in interdomain distances play a role only for peaks corresponding to crystallographic directions having approximate continuation across the twin planes (like [111]) and increase these peaks intensity. }
 \includegraphics[width=0.8\columnwidth]{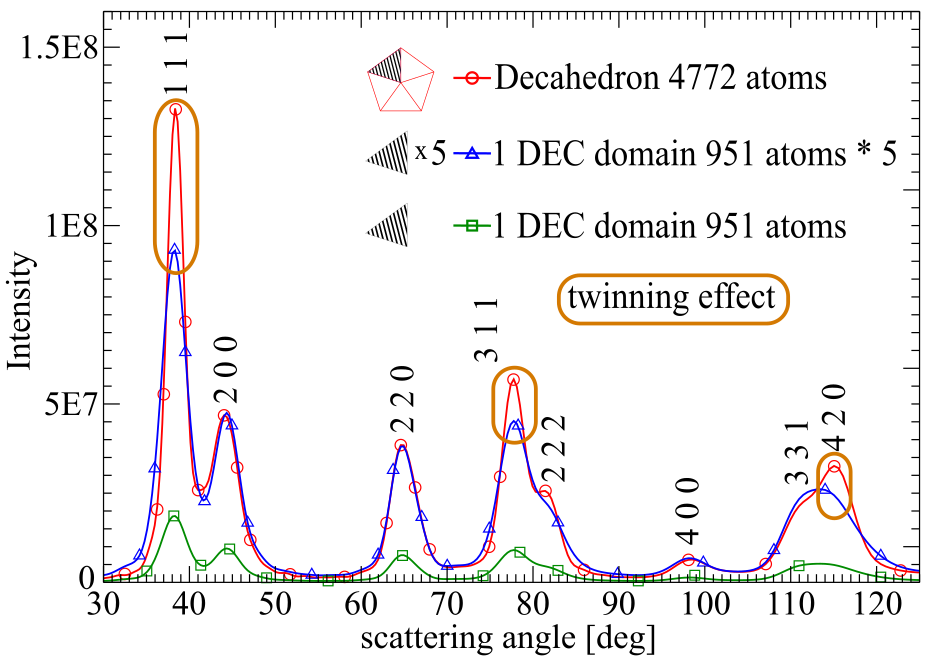}
 % \label{fgr:Fig14}
\end{figure}

\begin{figure}
 \centering
 \caption{Map of 'volumetric' strain (b) as given by the {\it{Ovito}} program \cite{Stukowski} for a single domain of a decahedron cut off around the 5-fold axis of an icosahedron (a). Half of the domain lays within the interior of icosahedron and is subjected to different relaxation forces. Appearance of contraction of distances in one part and of expansion in another part causes a splitting of the 220 reflection (c).}
 \includegraphics[width=0.8\columnwidth]{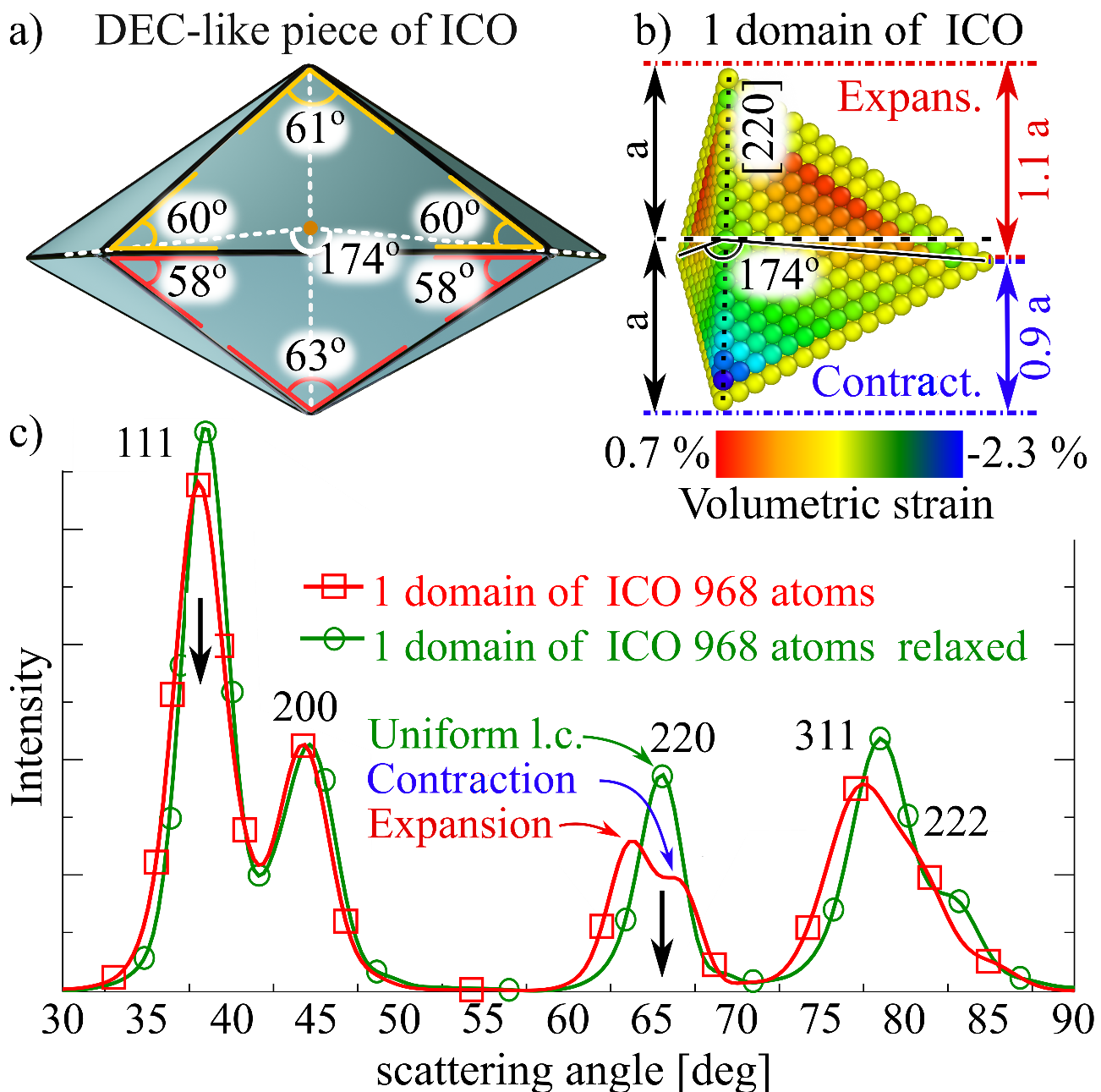}
 % \label{fgr:Fig15}
\end{figure}

\begin{figure}
 \centering
 \caption{111 and 200 peaks of diffraction patterns (Cu $K_{\alpha}$ radiation) for a range of Marks' decahedra. Intensity is divided by the number of atoms. The numbers of model atoms are given in the legend. For comparison added are diffraction patterns calculated for the proposed model of randomly multitwinned (MT) cluster consisting of 11158 atoms (orange curve), and of 5220 atoms (magenta curve), cut out of the model in the inset. In spite of the different model sizes, the peaks broadening is similar, driven by domain sizes and strain. The model MT structures have 111 peak width, distance between the peaks and their height ratio variable with the method parameters.}
 \includegraphics[width=0.8\columnwidth]{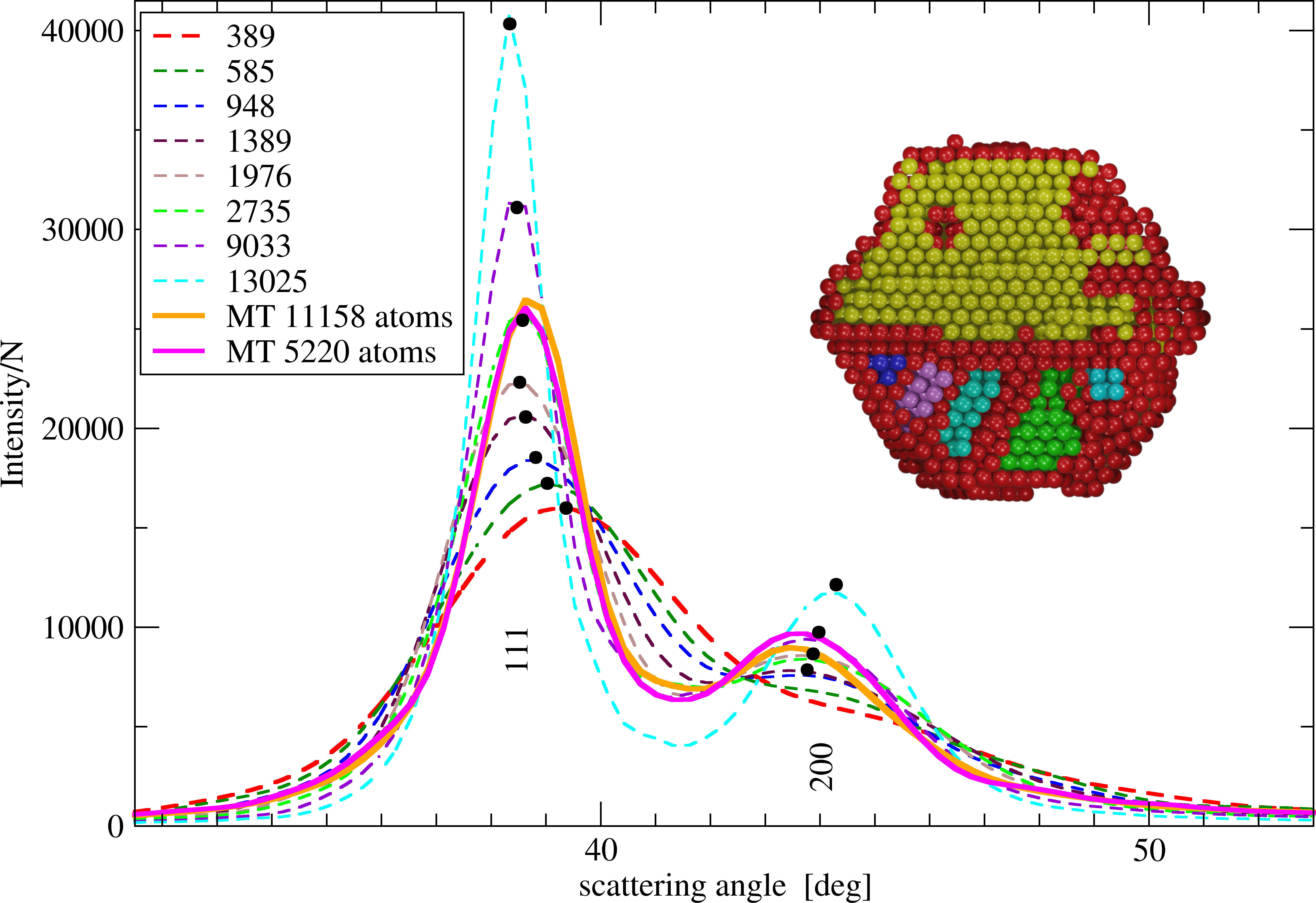}
 % \label{fgr:Fig16}
\end{figure}

\begin{figure}
 \centering
 \caption{Calculated diffraction patterns (Cu $K_{\alpha}$ radiation) for a range of icosahedra. Intensity is divided by the number of atoms. The numbers of model atoms are given in the legend.}
 \includegraphics[width=0.8\columnwidth]{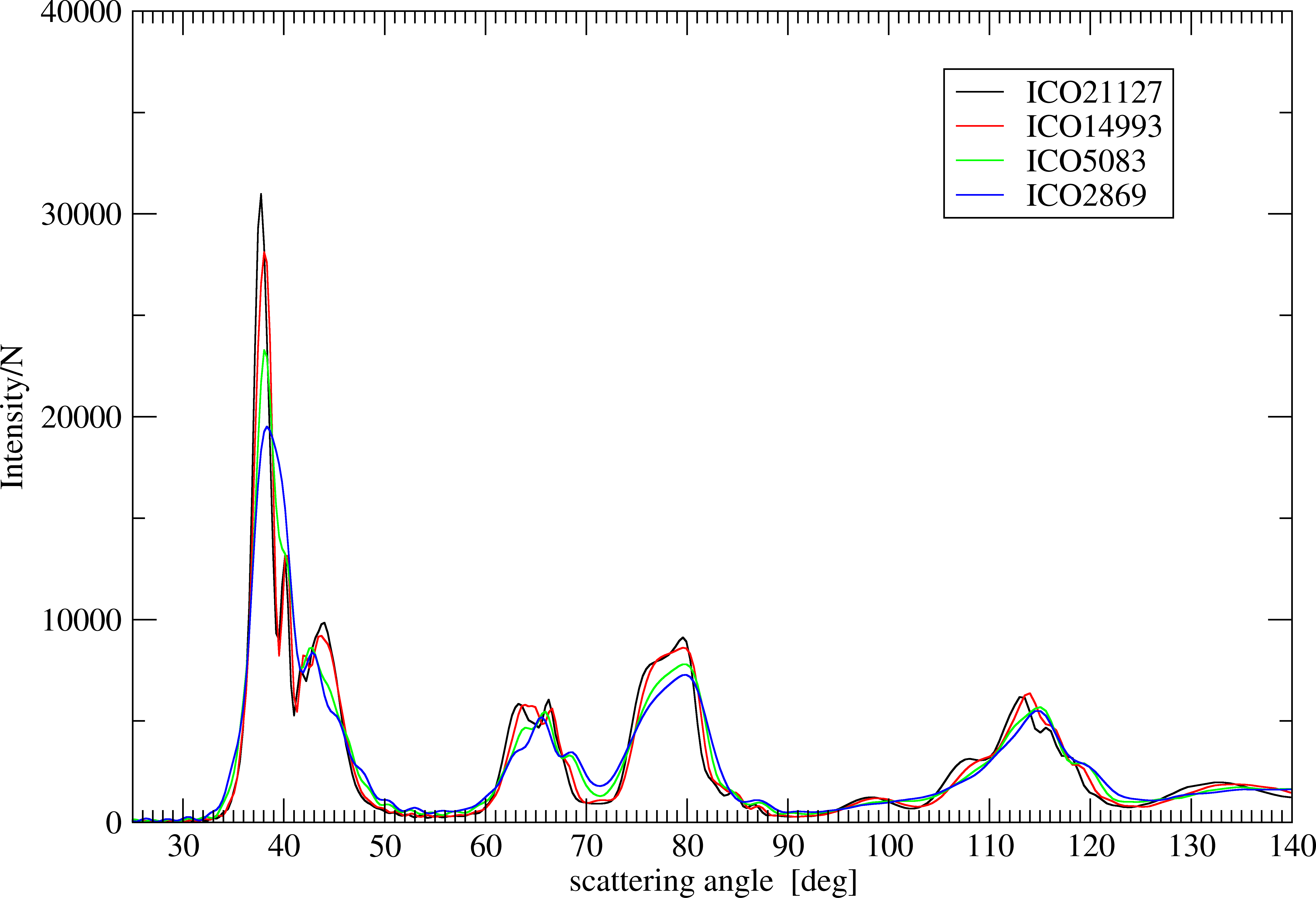}
 % \label{fgr:Fig17}
\end{figure}

\begin{figure}
 \centering
 \caption{Diffraction pattern (Cu $K_{\alpha}$ radiation) for a relaxed cubooctahedron consisting of 2869 atoms. The total intensity is compared to contributions from bulk-bulk, bulk-surface and surface-surface distances. }
 \includegraphics[width=0.8\columnwidth]{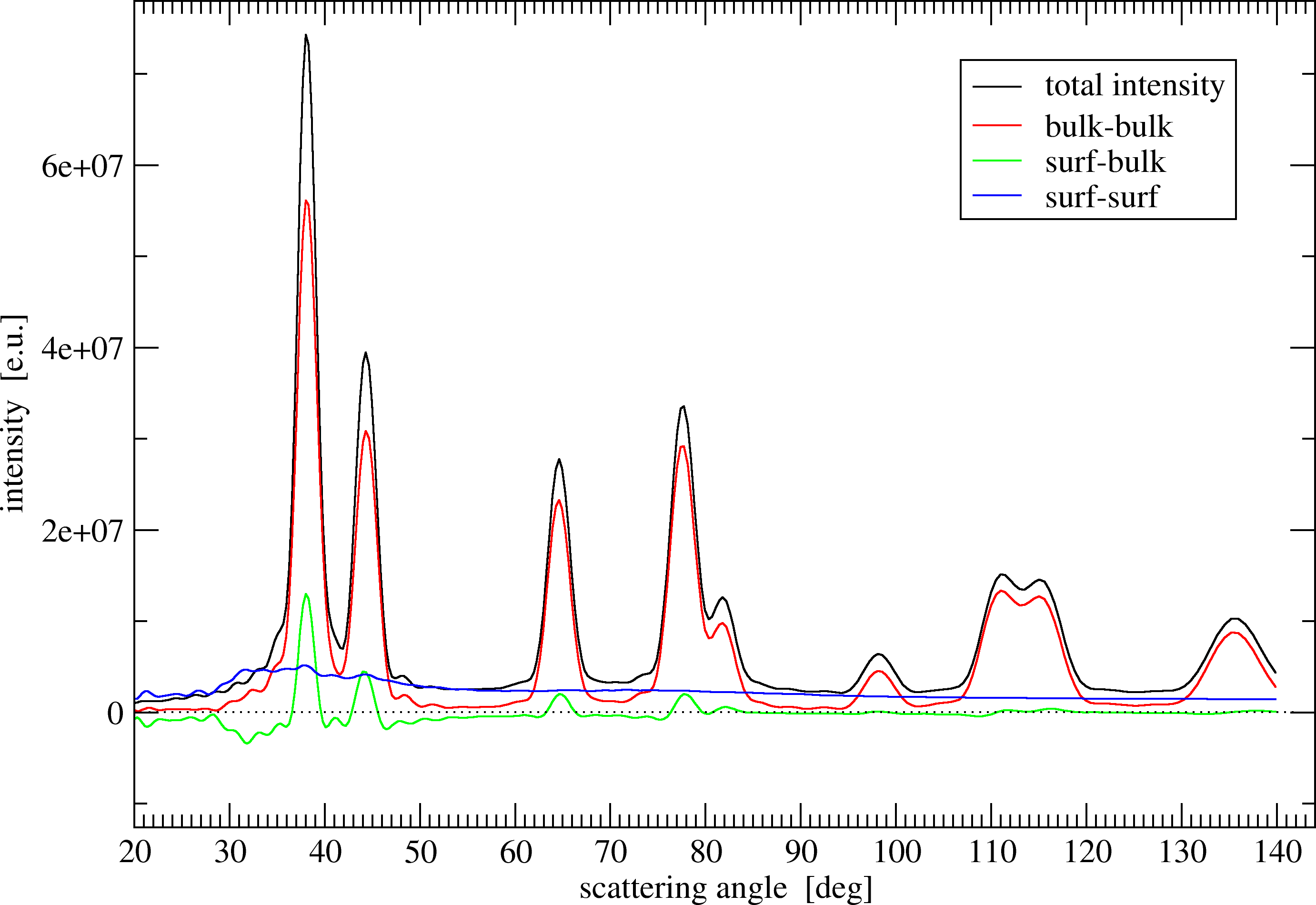}
 % \label{fgr:Fig18}
\end{figure}

     %-------------------------------------------------------------------------
     % TABLES AND FIGURES SHOULD BE INSERTED AFTER THE MAIN BODY OF THE TEXT
     %-------------------------------------------------------------------------
     % Simple tables should use the tabular environment according to this
     % model

     % Postscript figures can be included with multiple figure blocks
 %\begin{figure}
 % \centering
 % \caption{Representative in situ patterns of the 10\% Pt on silica sample in He after reduction in $H_{2}$  and after long exposure to CO and NO. The 
 % maximum at 23 deg corresponds to amorphous silica.}
 % \includegraphics[width=0.8\columnwidth]{Fig1}
 %\end{figure}    

\end{document}